\documentclass[reprint,aps,prl,superscriptaddress,nolongbibliography,amsmath,amssymb,floatfix]{revtex4-2}

\makeatletter
\providecommand\abx@aux@refcontext[1]{}
\providecommand\abx@aux@cite[2]{}
\providecommand\abx@aux@segm[3]{}
\providecommand\abx@aux@defaultrefcontext[3]{}
\providecommand\abx@aux@defaultlabelprefix[2]{}
\providecommand\abx@aux@read@bbl@mdfivesum[1]{}
\providecommand\abx@aux@read@bblrerun{}
\providecommand\blx@tocontentsinit[1]{}
\makeatother

\usepackage{graphicx}
\usepackage[colorlinks,linkcolor=blue,citecolor=blue,urlcolor=blue]{hyperref}

\graphicspath{{figures2/}{../figures2/}{Latex/figures2/}}

\newcommand{\runin}[1]{\textit{#1}---}

\begin{document}

\title{Operational Thresholds for Long-Range Entanglement in Quantum Networks}

\author{Arn Caesar}
\email[Contact author: ]{arn.caesar@me.com}
\affiliation{Blackett Laboratory and Centre for Complexity Science, Imperial College London, London SW7 2AZ, United Kingdom}

\author{Xinqi Hu}
\affiliation{Blackett Laboratory and Centre for Complexity Science, Imperial College London, London SW7 2AZ, United Kingdom}
\affiliation{School of Mathematical Sciences, Jiangsu University, Zhenjiang, Jiangsu 212013, China}

\author{Kim Christensen}
\affiliation{Blackett Laboratory and Centre for Complexity Science, Imperial College London, London SW7 2AZ, United Kingdom}

\begin{abstract}
Quantum networks distribute entanglement through probabilistic local operations, yet concurrence percolation assumes perfect operations. We derive an exact survival recursion on $(U,V)$-flower networks. For the $(2,2)$-flower, the operational threshold is $\eta_p=2-\tfrac{32}{27}\eta_s$. Here $\eta_s$ and $\eta_p$ denote series and parallel success probabilities. Survival jumps discontinuously at this threshold. A second threshold $c_{\mathrm{th}}(\eta_s,\eta_p)$ governs initial link quality. Above both, connections either fail or become Bell pairs. Both thresholds are key design constraints for scalable quantum networks.
\end{abstract}

\maketitle

\runin{Introduction}The prospect of a quantum internet \cite{doi:10.1126/science.aam9288,Kimble:2008wcw,Chen_2021,RevModPhys.95.045006} is driving quantum communication toward larger and more complex scales, where entanglement transmission \cite{Nielsen_Chuang_2010} becomes a collective, statistical problem \cite{Meng_2025,e1c8916eded64ff79fb79c5ce3ccc7b4}. Since distant nodes do not share entanglement directly, long-range entanglement must be assembled from short entangled links using local operations: entanglement swapping extends entanglement along a path \cite{PhysRevLett.71.4287,PhysRevLett.117.240503,doi:https://doi.org/10.1002/9781118648919.ch10}, while purification, concentration, and distillation combine parallel resources into stronger ones \cite{doi:10.1126/science.aan0070,Yan_2023,PhysRevLett.126.010503,DENG201746}. Establishing a useful end-to-end connection then depends on operational reliability, initial link strength, and network topology \cite{KUMAR2025104366,doi:https://doi.org/10.1002/9781118648919.ch6,doi:https://doi.org/10.1002/9781118648919.ch8}.

\begin{figure*}[t]
    \centering
    \includegraphics{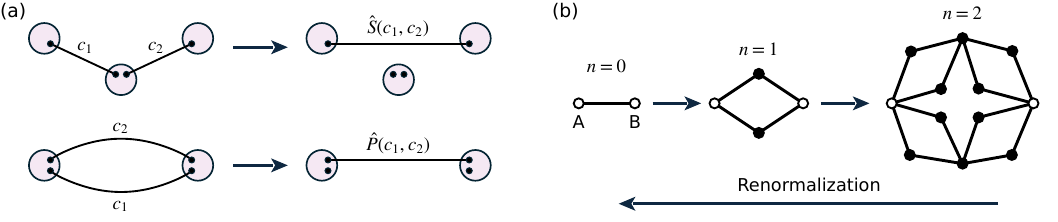}
    \caption{\looseness=-1 \textbf{Concurrence model of a quantum network and the hierarchical \((2,2)\)-flower.}
    (a)~Each link carries a concurrence \(c\in[0,1]\), with qubits shown as solid black dots inside nodes. Series combination (entanglement swapping) produces a weaker link of concurrence \(\hat{S}(c_1,c_2)\le\min(c_1,c_2)\), while parallel combination (purification--concentration) produces a stronger link of concurrence \(\hat{P}(c_1,c_2)\ge\max(c_1,c_2)\).
    (b)~The \((2,2)\)-flower is constructed by iteratively replacing each link with two parallel paths of length two, starting at generation \(n=0\). Boundary nodes A and B (open circles) persist across generations. Renormalization reverses this process, progressively reducing the network to a single effective link between A and B.}
    \label{fig:model}
\end{figure*}

Concurrence percolation \cite{PhysRevLett.126.170501} provides a framework for this collective problem. Each link is modeled as a partially entangled pure state with concurrence \(c\in[0,1]\) \cite{Wootters:2001ceg,PhysRevResearch.4.023059}, and local operations are represented by deterministic series and parallel rules \cite{PhysRevLett.126.170501,PhysRevResearch.5.013225}. On series--parallel reducible networks, the theory predicts sharp thresholds in initial concurrence below which long-range entanglement is impossible, and shows that quantum networks (QNs) are more strongly connected than predicted by classical percolation \cite{PhysRevLett.126.170501,e1c8916eded64ff79fb79c5ce3ccc7b4,doi:10.1126/sciadv.adt2404}. Hierarchical scale-free networks such as the \((U,V)\) flowers \cite{IKEDA2020109837,PhysRevE.75.061102} studied here are particularly valuable, as their recursive structure allows critical properties to be determined exactly \cite{doi:10.1126/sciadv.adt2404}.

These results, however, assume that every local operation succeeds with certainty. In practice, entanglement swapping and purification are probabilistic \cite{Nielsen_Chuang_2010,PhysRevLett.81.5932}, and entanglement degrades through environmental interactions \cite{YuESD2009,QuantEntDyn2001,almeida_SDE_2007,YuDephasingDecoherence2003,e25010157,9dw1-y4bl}. Stochastic connectivity is well established in classical network theory \cite{OLLE1997497,ATTAL20121545}, but its consequences for concurrence-based entanglement transmission remain unexplored. Two fundamental questions arise: how reliable must local operations be for long-range entanglement to survive, and how strong must the initial links be for the surviving connection to remain useful?

We address both questions by extending concurrence percolation to probabilistic operations. Series (entanglement swapping) and parallel (purification, concentration) operations succeed with probabilities \(\eta_s\) and \(\eta_p\), respectively, producing the ideal concurrence output upon success and destroying the effective link upon failure. We study this model on \((U,V)\) flowers, focusing on the \((2,2)\)-flower, whose recursive structure permits an exact renormalization procedure \cite{christensen2005complexity,e1c8916eded64ff79fb79c5ce3ccc7b4}.

Our central finding is a two-threshold structure for long-range entanglement. First, an exact recursion for the survival probability of the end-to-end connection yields a closed-form operational phase boundary, \(\eta_p = 2-\tfrac{32}{27}\eta_s\) for the \((2,2)\)-flower. Below this boundary, long-range entanglement vanishes regardless of initial link quality; at the boundary, the asymptotic survival probability jumps discontinuously from zero to a finite value, \(3/(4\eta_s)\). Second, above this boundary, a threshold in initial concurrence remains. The nonlinear parallel rule prevents an exact recursion for the surviving concurrence, so we determine this threshold by numerically propagating its probability distribution. Above both thresholds, the large-network distribution becomes binary: complete failure or a Bell pair. Thus, operational noise produces an all-or-nothing outcome with probabilities determined exactly by the survival theory. These results place operational success rates alongside initial link quality as a key design constraint for scalable QNs \cite{PhysRevLett.81.5932,RevModPhys.95.045006}.

\runin{Model}In the QN model, each link represents a bipartite pure state shared between two qubits at its end nodes [Fig.~\ref{fig:model}(a)] \cite{PhysRevResearch.4.023059}. For Schmidt coefficients \(\lambda_1\) and \(\lambda_2\), its concurrence is \(c = 2\sqrt{\lambda_1\lambda_2}\in[0,1]\) \cite{Wootters:2001ceg}, where \(c=0\) means no entanglement and \(c=1\) a Bell pair (Supplemental Material \cite{SMarXiv}, hereafter SM, Sec.~S1). The QN is thereby reduced to a weighted network, with every link initially carrying concurrence \(c_0\).

Long-range entanglement between two distant nodes, Alice (A) and Bob (B), is built by repeatedly combining links in series (entanglement swapping) or in parallel (purification--concentration) [Fig.~\ref{fig:model}(a)]. When successful, these operations produce the concurrence outputs \(\hat{S}(c_1,c_2)\le\min(c_1,c_2)\) and \(\hat{P}(c_1,c_2)\ge\max(c_1,c_2)\) of ideal concurrence percolation \cite{PhysRevLett.126.170501} (SM, Sec.~S1). Our framework relies only on these outputs, not on protocol- or hardware-specific mechanisms \cite{doi:10.1126/science.aam9288,Nielsen_Chuang_2010}.

We study the \((U,V)\) flowers \cite{IKEDA2020109837,PhysRevE.75.061102}, hierarchical networks grown by iteratively replacing each link with two parallel paths of lengths \(2\le U\le V\) [Fig.~\ref{fig:model}(b)]. The generation \(n\) network contains \((U+V)^n\) links (measure of network size), while the shortest path between the boundary nodes A and B grows as \(L=U^n\). Crucially, the recursive structure admits an exact renormalization: each step coarse grains every elementary motif into a single link, \(\hat{P}\big(\hat{S}(c_1,c_2),\hat{S}(c_3,c_4)\big)\) for a \((2,2)\)-flower motif with link concurrences \(c_1,c_2\) and \(c_3,c_4\), so that \(n\) steps reduce a generation \(n\) network to a single effective link between A and B \cite{e1c8916eded64ff79fb79c5ce3ccc7b4,christensen2005complexity}.

\begin{figure}[h!]
    \centering
    \includegraphics{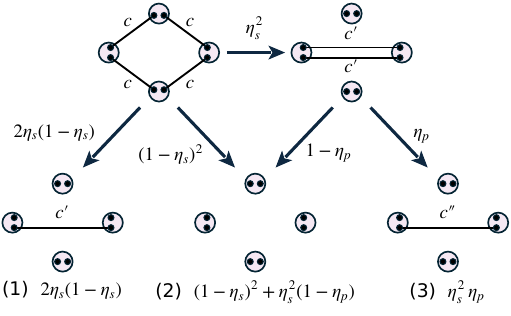}
    \caption{\textbf{Stochastic renormalization of a \((2,2)\)-flower motif} with concurrence \(c\) on each link. Each series operation succeeds independently with probability \(\eta_s\), producing concurrence \(c'=\hat{S}(c,c)\le c\). If both succeed, the parallel operation succeeds with probability \(\eta_p\), yielding \(c''=\hat{P}(c',c')\ge c'\). Failure destroys the affected link. Three outcomes are possible, labeled by their total probabilities: (1)~one surviving path of concurrence \(c'\); (2)~complete disconnection; and (3)~a parallel-combined link of concurrence \(c''\). Thus, the end-to-end concurrence \(X_n\) is a random variable.}
    \label{fig:outcomes}
\end{figure}

Probabilistic operations make each renormalization step random: a single step of the \((2,2)\)-flower motif may yield a single surviving path, no connection at all, or a parallel-combined link (Fig.~\ref{fig:outcomes}). The end-to-end concurrence of a generation \(n\) network, \(X_n\in[0,1]\), is therefore a random variable, with \(3^{4^{n-1}}\) possible operational histories for the \((2,2)\)-flower, one of three outcomes for each of its \(4^{n-1}\) elementary motifs. Its distribution encodes the two central questions of this work: does long-range entanglement survive, and if so, is the surviving connection sufficiently strong to be useful?

\runin{Exact survival theory}The most basic observable is the survival probability,
\begin{equation}
    S_n = \mathbb{P}(X_n>0),
    \label{eq:Sn_definition}
\end{equation}
the probability that A and B remain connected by nonzero entanglement after a generation \(n\) network is reduced to a single effective link. Because the \((U,V)\) flowers are self-similar, \(S_n\) obeys an exact renormalization-group (RG) recursion: a generation \(n+1\) network is an elementary motif whose links are generation \(n\) subnetworks, each independently surviving with probability \(S_n\). Enumerating the success--failure outcomes of one motif (Fig.~\ref{fig:outcomes}) gives, for the \((2,2)\)-flower,
\begin{equation}
    S_{n+1}=\mathcal{R}(S_n)=2\eta_s S_n^2-(2-\eta_p)\,\eta_s^2 S_n^4,\quad S_0=1.
    \label{eq:survival_recursion}
\end{equation}
For the general \((U,V)\)-flower derivation, see SM, Sec.~S2. Equation~\eqref{eq:survival_recursion} exactly resums all \(3^{4^{n-1}}\) operational histories through the single scalar \(S_n\).

\begin{figure}[t]
    \centering
    \includegraphics{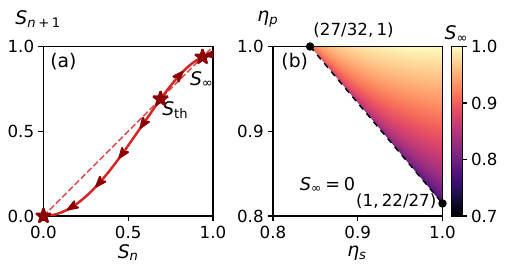}
    \caption{\textbf{Exact survival theory and operational phase diagram.}
    (a)~Survival recursion \(S_{n+1}=\mathcal{R}(S_n)\) (solid curve), Eq.~\eqref{eq:survival_recursion}, at \(\eta_s=\eta_p=0.95\), and the identity (dashed line). Stars (\(\bigstar\)) mark fixed points and arrows indicate RG flow. The fixed points \(S^*=0\) and \(S^*=S_\infty\) are stable, while \(S^*=S_{\mathrm{th}}\) is unstable. Lowering \(\eta_s\) and \(\eta_p\) causes \(S_\infty\) and \(S_{\mathrm{th}}\) to merge, defining the operational phase boundary in (b).
    (b)~Asymptotic survival probability \(S_\infty\) versus \((\eta_s,\eta_p)\) for \(c_0=1\). The dashed line marks the exact phase boundary, Eq.~\eqref{eq:phase_boundary}, below which \(S_\infty=0\). On the boundary \(S_\infty=3/(4\eta_s)\), giving the discontinuous jump in survival probability.}
    \label{fig:survival}
\end{figure}

The large-network behavior follows from the fixed points \(S^*=\mathcal{R}(S^*)\) [Fig.~\ref{fig:survival}(a)]. The map has a stable fixed point at \(S^*=0\) (complete failure of long-range entanglement), and, for sufficiently reliable operations, a stable nonzero fixed point \(S^*=S_\infty=\lim_{n\to\infty}S_n\) together with an unstable fixed point \(S^*=S_{\mathrm{th}}\) separating the two basins of attraction. Notably, \(S_\infty<1\) throughout: operational failures always leave a finite probability that the connection is lost, however large the network. Since the recursion is independent of \(c_0>0\), and every surviving realization is a Bell pair when \(c_0=1\), the survival theory completely determines entanglement transmission for perfect initial links.

\runin{Exact operational phase diagram}Because Eq.~\eqref{eq:survival_recursion} depends only on \(\eta_s\) and \(\eta_p\), its fixed-point structure partitions the operational parameter space exactly. Figure~\ref{fig:survival}(b), the central result of this work, shows the asymptotic survival probability \(S_\infty\) across the \((\eta_s,\eta_p)\) plane. The two phases, in which long-range entanglement is asymptotically impossible (\(S_\infty=0\)) or possible with finite probability (\(0<S_\infty\le1\)), are separated by the boundary at which the stable and unstable fixed points merge, \(S_{\mathrm{th}}=S_\infty\). Solving this tangency condition analytically (SM, Sec.~S2) yields an exact phase boundary for the \((2,2)\)-flower:
\begin{equation}
    \eta_p = 2-\tfrac{32}{27}\,\eta_s,
    \qquad \tfrac{27}{32}\le\eta_s\le 1.
    \label{eq:phase_boundary}
\end{equation}
For equally reliable operations, the critical success probability is \(\eta_s=\eta_p=54/59\approx0.92\). Even with Bell pairs initially (\(c_0=1\)), operations must succeed more than nine times out of ten.

The transition across the boundary is discontinuous. At the phase boundary, \(S_\infty=3/(4\eta_s)\) (SM, Sec.~S2), whereas below it, \(S_\infty=0\). Thus the survival probability jumps from zero to at least \(3/4\), unlike the continuous transition of ordinary second-order percolation \cite{christensen2005complexity,PhysRevE.75.061102}. The same picture persists for general \((U,V)\) flowers, whose longer branches shrink the surviving phase. Close to the operational phase boundary, finite networks can exhibit substantial survival probabilities even below threshold, owing to a slow approach to the large-network limit (SM, Sec.~S2).

\runin{Strength of the surviving connection}Survival alone does not guarantee useful entanglement \cite{YuESD2009,KUMAR2025104366}: for imperfect initial links, \(c_0<1\), repeated series combination weakens the connection exponentially, while parallel combination strengthens it. To quantify the strength of the surviving connection, we introduce the conditional mean concurrence,
\begin{equation}
    Y_n = \mathbb{E}[X_n \mid X_n>0],
    \qquad
    \mathbb{E}[X_n] = S_n Y_n,
    \label{eq:Yn_definition}
\end{equation}
the average strength of the connection in realizations where it survives; the decomposition separates how often a connection exists, \(S_n\) (known exactly), from how strong it is when it does, \(Y_n\).

\begin{figure}[t]
    \centering
    \includegraphics{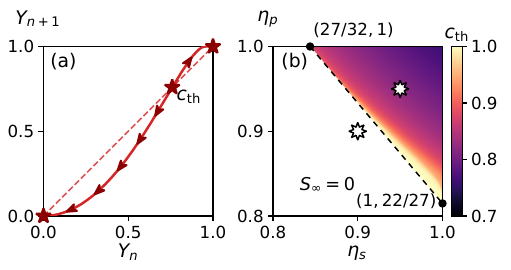}
    \caption{\textbf{Concurrence threshold in the ideal limit and under probabilistic operations.}
    (a)~Deterministic concurrence recursion, Eq.~\eqref{eq:ideal_recursion}, for \(\eta_s=\eta_p=1\), where \(S_n=1\) and \(X_n=Y_n\), using the conventions of Fig.~\ref{fig:survival}(a). The stable fixed points \(Y^*=0\) and \(Y^*=1\) are separated by the unstable fixed point \(Y^*=c_{\mathrm{th}}=0.75947\ldots\) \cite{doi:10.1126/sciadv.adt2404}.
    (b)~Concurrence threshold \(c_{\mathrm{th}}\), obtained by propagating the distribution of \(X_n\) to generation \(n=50\) and locating the transition by bisection, using the color scale of Fig.~\ref{fig:survival}(b). The dashed line marks the operational phase boundary, Eq.~\eqref{eq:phase_boundary}, below which \(S_\infty=0\). Open stars mark \((\eta_s,\eta_p)=(0.90,0.90)\) and \((0.95,0.95)\), used in Fig.~\ref{fig:pmf}. The threshold decreases with increasing operational reliability, reaching the ideal value of (a) at \((\eta_s,\eta_p)=(1,1)\).}
    \label{fig:concurrence}
\end{figure}

With perfect operations, \(\eta_s=\eta_p=1\), every operation succeeds, so \(S_n=1\), and \(X_n=Y_n\) obeys the exact recursion of deterministic concurrence percolation \cite{doi:10.1126/sciadv.adt2404},
\begin{equation}
    Y_{n+1} = \hat{P}\big(\hat{S}(Y_n,Y_n),\,\hat{S}(Y_n,Y_n)\big),
    \qquad Y_0=c_0.
    \label{eq:ideal_recursion}
\end{equation}
Its stable fixed points \(Y^*=0\) and \(Y^*=1\) are separated by an unstable fixed point at the concurrence threshold \(c_{\mathrm{th}}=0.75947\ldots\) [Fig.~\ref{fig:concurrence}(a)], known to arbitrary precision \cite{PhysRevLett.126.170501,doi:10.1126/sciadv.adt2404} (SM, Sec.~S4): for \(c_0<c_{\mathrm{th}}\), the end-to-end entanglement vanishes in the large-network limit whereas for \(c_0>c_{\mathrm{th}}\) it approaches a Bell pair.

Under probabilistic operations \(X_n\) becomes a random variable, and the nonlinear parallel combination prevents the exact recursion of Eq.~\eqref{eq:ideal_recursion}. We therefore numerically propagate the distribution of \(X_n\) to determine the concurrence threshold [Fig.~\ref{fig:concurrence}(b) and SM, Sec.~S3].

\runin{Concurrence threshold}The transition between vanishing and near-perfect concurrence above the operational threshold defines the concurrence threshold \(c_{\mathrm{th}}(\eta_s,\eta_p)\). At large \(n\), the probability of obtaining a near-perfect end-to-end connection approaches a step function in the initial concurrence \(c_0\), rising from zero to \(S_\infty\) at \(c_{\mathrm{th}}\).

Figure~\ref{fig:concurrence}(b) maps \(c_{\mathrm{th}}\) at \(n=50\): the threshold decreases smoothly as operations become more reliable, reaching the ideal value \(c_{\mathrm{th}}=0.75947\ldots\) at \(\eta_s=\eta_p=1\), and increases toward the operational phase boundary, varying continuously along it.

\runin{Distribution of the end-to-end concurrence}Our numerical method propagates the distribution of \(X_n\), including the probability of complete failure, \(\mathbb{P}(X_n=0)\), through successive renormalization steps (SM, Sec.~S5). The method retains the distribution without sampling noise and reaches generation \(n=50\), with \((U+V)^n\approx10^{30}\) links and \(L=U^n\approx10^{15}\), while maintaining negligible numerical error (SM, Sec.~S6).

\begin{figure*}[t]
    \centering
    \includegraphics{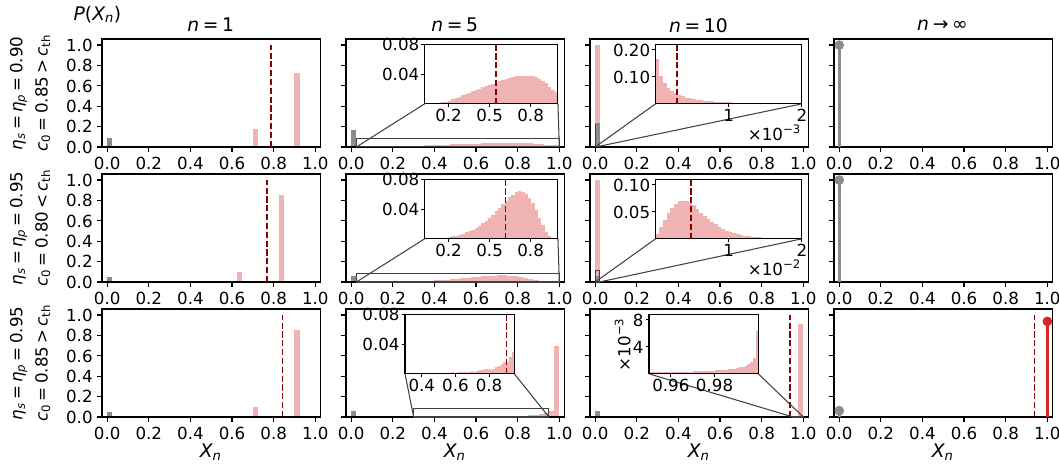}
    \caption{\looseness=-1 \textbf{Evolution of the end-to-end concurrence distribution across three regimes.}
    Probability distribution of \(X_n\) at generations \(n=1\), \(5\), and \(10\) and \(n\to\infty\), for identical initial links. Top row: below the operational threshold, \(\eta_s=\eta_p=0.90\) with \(c_0=0.85\). Middle row: above the operational but below the concurrence threshold, \(\eta_s=\eta_p=0.95\) with \(c_0=0.80<c_{\mathrm{th}}(\eta_s,\eta_p)\approx0.82\). Bottom row: above both thresholds, \(\eta_s=\eta_p=0.95\) with \(c_0=0.85>c_{\mathrm{th}}(\eta_s,\eta_p)\approx0.82\). Finite-generation distributions use bins of width \(0.025\); dashed lines mark the mean \(\mathbb{E}[X_n]\), shown in the inset when within its range. Gray denotes complete failure, \(\mathbb{P}(X_n=0)=1-S_n\), and red denotes surviving connections, \(X_n>0\). Insets magnify the outlined regions. In the large-network limit, delta functions are represented by stems with dots, indicating their probability weights. The first two regimes converge to \(\delta(X)\), while above both thresholds the distribution becomes binary, \((1-S_\infty)\,\delta(X)+S_\infty\,\delta(X-1)\), with \(S_\infty\simeq0.94\); thus, Bell pairs account for approximately \(94\%\) of the probability in the bottom row.}
    \label{fig:pmf}
\end{figure*}

Figure~\ref{fig:pmf} shows the evolution of the end-to-end concurrence distribution in three regimes: below the operational threshold, above the operational but below the concurrence threshold, and above both thresholds. At \(n=1\), each distribution exhibits the three outcomes of Fig.~\ref{fig:outcomes}: complete failure, \(X_1=0\), and the two surviving concurrences \(c'\) and \(c''\); as \(n\) increases, the number of operational histories grows rapidly and the distributions reveal the distinct behavior of each regime.

Below the operational threshold (\(\eta_s=\eta_p=0.90\), \(c_0=0.85\)), a continuum of weakened surviving connections forms alongside the probability mass at \(X_n=0\). With increasing generation, their concurrence decreases and the probability of complete failure grows, until \(S_n\to0\) and the distribution collapses to \(X_n=0\) as \(n\to\infty\).

Above the operational but below the concurrence threshold (\(\eta_s=\eta_p=0.95\), \(c_0=0.80\)), the survival probability approaches a nonzero limit, \(S_\infty>0\), but the initial links are too weak to sustain useful long-range entanglement. The concurrence of surviving connections approaches zero, and the distribution collapses to \(X_n=0\) as \(n\to\infty\).

Above both thresholds (\(\eta_s=\eta_p=0.95\), \(c_0=0.85\)), the probability of complete failure approaches \(1-S_\infty\), while the remaining probability concentrates near \(X_n=1\). In the large-network limit, the distribution becomes binary: complete failure with probability \(1-S_\infty\), or a Bell pair with probability \(S_\infty\simeq0.94\). Thus, \(\mathbb{E}[X_n]\to S_\infty\), as confirmed numerically (SM, Fig.~S4).

The final column of Fig.~\ref{fig:pmf} thus summarizes the central result: below the operational threshold, long-range connections disappear; above the operational but below the concurrence threshold, they survive but their concurrence vanishes; and above both thresholds, transmission is all-or-nothing, with a success probability determined exactly by the survival theory.

\runin{Discussion}Our results reveal a two-threshold structure governing long-range entanglement in quantum networks with probabilistic local operations: the operational threshold determines whether an end-to-end connection can survive at all, and the concurrence threshold whether it is sufficiently entangled to be useful. Together, they quantify two complementary requirements for scalable entanglement transmission: sufficiently reliable operations and sufficiently strong initial links.

Since operations must succeed with probability \(\eta_s=\eta_p=54/59\simeq0.92\) even for the favorable \((2,2)\)-flower topology with perfect initial links, operational reliability is not merely an implementation detail but a fundamental constraint alongside link quality. This requirement grows as series paths lengthen, rising to \(0.95\) and \(0.97\) for the \((2,3)\)- and \((2,4)\)-flowers (SM, Sec.~S2), and connects directly to ongoing efforts to raise the success probabilities and fidelities of quantum hardware \cite{GHASEMI2020126728,LI2026100432,paetznick2024demonstration,Heeres_2017}. Above the operational threshold, these two requirements can partly compensate for one another, as quantified in Fig.~\ref{fig:concurrence}(b): stronger initial links tolerate less reliable operations, and vice versa. Parallel operations further highlight the value of network redundancy, as alternative paths reinforce imperfect entanglement \cite{doi:https://doi.org/10.1002/9781118648919.ch10,Yan_2023,PhysRevLett.126.010503}.

\runin{Conclusion and outlook}Probabilistic local operations fundamentally change the concurrence-percolation problem: long-range entanglement is controlled not only by initial link quality but also by a distinct threshold in operational reliability, and above both thresholds the connection either fails completely or approaches a Bell pair, with a probability set by the survival theory.

The \((U,V)\)-flower is a deliberately simplified setting in which this collective effect can be treated analytically. Realistic networks need not be series--parallel reducible \cite{Tian2024QuantumNF}, and our treatment neglects heterogeneous links, memory decoherence, and time-dependent noise \cite{YuESD2009,QuantEntDyn2001,YuDephasingDecoherence2003,e25010157,9dw1-y4bl}, mixed-state entanglement \cite{Wang:2026snq,Felicetti_2002}, and adaptive protocols such as routing and repeated attempts \cite{doi:10.1126/science.aam9288,KUMAR2025104366}. Extending the framework to these effects is a natural next step; mixed-state generalizations in particular \cite{Wang:2026snq} could test whether the two-threshold structure persists beyond pure-state concurrence percolation.

Despite these simplifications, our results reveal a fundamental limitation of quantum networks: even perfect entanglement between neighboring nodes cannot sustain long-range entanglement if local operations are insufficiently reliable.

\begin{acknowledgments}
\runin{Acknowledgments}We thank Tao Baker for sharing relevant papers and for insightful discussions.
\end{acknowledgments}

\bigskip
\noindent A.C.\ designed the research, performed the analytical derivations, developed the simulations, analyzed the results, and wrote the original draft. K.C.\ supervised the project and revised the manuscript. X.H.\ contributed to the interpretation of the results and revised the manuscript.

\bigskip
\noindent The authors declare no competing interests.

\bigskip
{\interlinepenalty=10000
\runin{Data availability}The data that support the findings of this article are contained in the article and its Supplemental Material \cite{SMarXiv}. The code used to generate the numerical results is openly available \cite{code}.\par}

\bibliography{reference}

\clearpage
\onecolumngrid
\setcounter{secnumdepth}{3}
\setcounter{section}{0}
\setcounter{equation}{0}
\setcounter{figure}{0}
\setcounter{table}{0}
\renewcommand{\thesection}{S\arabic{section}}
\renewcommand{\thesubsection}{\thesection.\arabic{subsection}}
\makeatletter
\renewcommand{\p@subsection}{}
\makeatother
\renewcommand{\theequation}{S\arabic{equation}}
\renewcommand{\thefigure}{S\arabic{figure}}
\renewcommand{\thetable}{S\arabic{table}}
\renewcommand{\theHsection}{S\arabic{section}}
\renewcommand{\theHsubsection}{S\arabic{section}.\arabic{subsection}}
\renewcommand{\theHequation}{S\arabic{equation}}
\renewcommand{\theHfigure}{S\arabic{figure}}
\renewcommand{\theHtable}{S\arabic{table}}
\renewcommand{\labelitemi}{\(\bullet\)}

\begin{center}
    {\large\bfseries Supplemental Material for ``Operational Thresholds for Long-Range Entanglement in Quantum Networks''\par}
    \vspace{0.8em}
    Arn Caesar, Xinqi Hu, and Kim Christensen
\end{center}

\vspace{0.5em}
\noindent Sections, equations, figures, and tables numbered with the prefix S refer to this Supplemental Material; those without it refer to the main text. Reference numbers refer to the reference list of the main text.

\section{Concurrence and the series and parallel rules}
\label{sec:S1}

Each link is a two-qubit pure state. In suitable local bases it has the Schmidt form
\begin{equation}
    |\psi\rangle=\sqrt{\lambda_1}\,|00\rangle+\sqrt{\lambda_2}\,|11\rangle,
    \qquad \lambda_1+\lambda_2=1,\quad \lambda_1\ge\lambda_2\ge0,
\end{equation}
and its concurrence is \(c=2\sqrt{\lambda_1\lambda_2}\) \cite{Wootters:2001ceg}, with \(c=1\) for a Bell pair and \(c=0\) for a product state. Inverting, the larger Schmidt coefficient is
\begin{equation}
    \lambda_1(c)=\tfrac{1}{2}\big(1+\sqrt{1-c^2}\big).
    \label{eq:lambda}
\end{equation}

When successful, the series and parallel operations give the concurrences of ideal concurrence percolation \cite{PhysRevLett.126.170501}:
\begin{align}
    \hat S(c_1,c_2)&=c_1c_2, \label{eq:S_rule}\\
    \hat P(c_1,c_2)&=2\sqrt{M(1-M)},\qquad M=\max\!\big\{\lambda_1(c_1)\,\lambda_1(c_2),\,\tfrac12\big\}. \label{eq:P_rule}
\end{align}
The series rule describes entanglement swapping. The parallel rule describes deterministic entanglement concentration: the two links together form a state whose largest Schmidt coefficient is \(\lambda_1(c_1)\lambda_1(c_2)\), and by Nielsen's majorization theorem the most entangled two-qubit state that can be obtained from it with certainty by local operations has largest Schmidt coefficient \(M\) \cite{Nielsen_Chuang_2010}. Series combination never strengthens a link, \(\hat S(c_1,c_2)\le\min(c_1,c_2)\), and parallel combination never weakens one, \(\hat P(c_1,c_2)\ge\max(c_1,c_2)\), with \(\hat P=1\) whenever \(\lambda_1(c_1)\lambda_1(c_2)\le\tfrac12\). The survival theory below uses only whether an operation succeeds, not these explicit forms.

\section{Exact survival theory}
\label{sec:S2}

\subsection{Survival recursion}

A generation \(n+1\) network is one motif whose links are generation \(n\) subnetworks. Each subnetwork reduces to a single effective link that is nonzero with probability \(S_n\), independently of the others. In the \((2,2)\)-flower, a branch of two subnetworks survives if both are nonzero and the series operation joining them succeeds,
\begin{equation}
    p_n=\eta_s S_n^2 .
\end{equation}
The motif survives if exactly one branch survives, with probability \(2p_n(1-p_n)\), or if both survive and the parallel operation succeeds, with probability \(\eta_p p_n^2\). Hence
\begin{equation}
    S_{n+1}=2p_n(1-p_n)+\eta_p p_n^2=2\eta_s S_n^2-(2-\eta_p)\,\eta_s^2S_n^4,\qquad S_0=1,
    \label{eq:survival22}
\end{equation}
which is Eq.~(2).

\begin{figure}[!tb]
    \centering
    \includegraphics{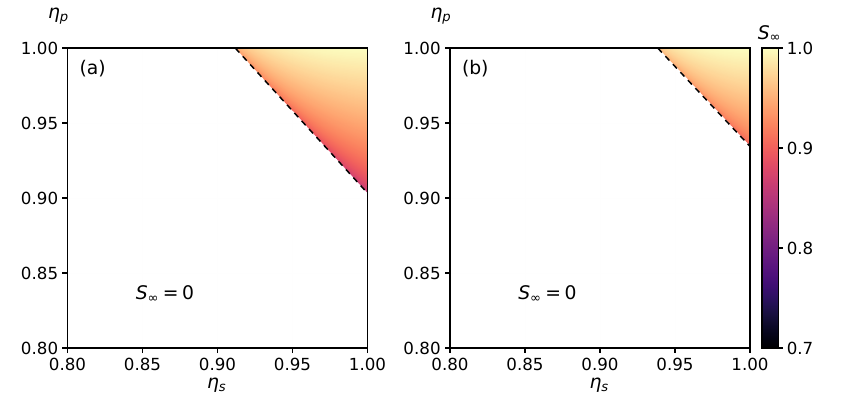}
    \caption{\textbf{Operational phase diagrams of longer flowers.} Asymptotic survival probability \(S_\infty\), the stable fixed point of Eq.~\eqref{eq:survivalUV}, for (a)~the \((2,3)\)-flower and (b)~the \((2,4)\)-flower, with the same conventions as Fig.~3(b). The surviving phase shrinks as the longer branch grows.}
    \label{fig:flowers}
\end{figure}

For a general \((U,V)\)-flower, a branch of length \(U\) contains \(U\) subnetworks joined by \(U-1\) series operations, so it survives with probability \(P_U=\eta_s^{U-1}S_n^U\), and likewise \(P_V=\eta_s^{V-1}S_n^V\). The same counting gives
\begin{equation}
    S_{n+1}=P_U+P_V-(2-\eta_p)\,P_UP_V,\qquad S_0=1 .
    \label{eq:survivalUV}
\end{equation}
The fixed-point structure is unchanged: \(S^*=0\) is stable and, for sufficiently reliable operations, a stable \(S_\infty\) and an unstable \(S_{\mathrm{th}}\) appear together. Longer branches need more successful operations, so the surviving phase shrinks (Fig.~\ref{fig:flowers}). For equal reliabilities, \(\eta_s=\eta_p\), the critical success probability rises from \(54/59\approx0.915\) for the \((2,2)\)-flower to \(0.954\) for the \((2,3)\)-flower and \(0.968\) for the \((2,4)\)-flower.

\subsection{Operational phase boundary}

Nonzero fixed points of Eq.~\eqref{eq:survival22} satisfy \(S_{n+1}/S_n=1\), that is
\begin{equation}
    h(S)\equiv 2\eta_s S-(2-\eta_p)\,\eta_s^2S^3=1 .
    \label{eq:fp}
\end{equation}
The cubic \(h\) has its maximum at \(S_m^2=2/[3(2-\eta_p)\eta_s]\), where \(h(S_m)=\tfrac43\eta_sS_m\). If \(h(S_m)>1\), Eq.~\eqref{eq:fp} has two roots, \(S_{\mathrm{th}}<S_\infty\); if \(h(S_m)<1\), it has none and \(S_n\to0\). The two roots merge when \(h(S_m)=1\), that is at
\begin{equation}
    S_m=\frac{3}{4\eta_s}.
\end{equation}
Substituting into the expression for \(S_m^2\) gives \(2-\eta_p=\tfrac{32}{27}\eta_s\), the boundary of Eq.~(3). Since \(\eta_p\le1\), the boundary requires \(\eta_s\ge27/32\), and for \(\eta_s=\eta_p\) it gives \(54/59\). On the boundary \(h(S)<1\) for every \(S\neq S_m\), so \(S_n\) decreases monotonically from \(S_0=1\) and converges to the merged fixed point, \(S_\infty=3/(4\eta_s)\).

\subsection{Critical slowing down}
\label{sec:csd}

With \(u_n=\eta_sS_n\), Eq.~\eqref{eq:survival22} becomes
\begin{equation}
    u_{n+1}=2u_n^2-\kappa u_n^4,\qquad \kappa=\frac{2-\eta_p}{\eta_s},
\end{equation}
so \(\eta_s\) and \(\eta_p\) enter the dynamics only through \(\kappa\). The nonzero fixed points merge at \(u=3/4\) when \(\kappa=32/27\). We measure the distance from the boundary by
\begin{equation}
    \epsilon=\frac{2-\eta_p}{\eta_s}-\frac{32}{27},
\end{equation}
which is positive below the boundary. At the merging point the map minus the identity has a double zero, with second derivative \(-4\), and its derivative with respect to \(\kappa\) is \(-(3/4)^4=-81/256\). To leading order, \(x_n=u_n-3/4\) therefore obeys
\begin{equation}
    x_{n+1}-x_n=-\tfrac{81}{256}\,\epsilon-2x_n^2 .
    \label{eq:normal}
\end{equation}

\begin{figure}[!tb]
    \centering
    \includegraphics{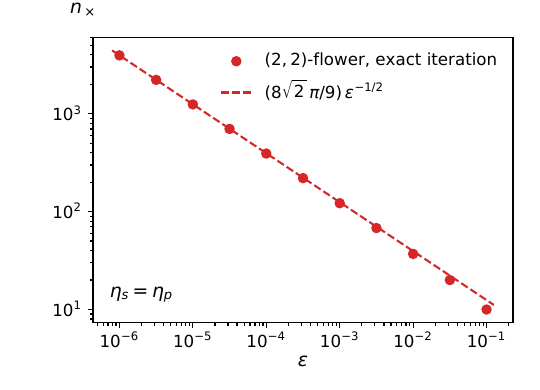}
    \caption{\textbf{Critical slowing down below the operational phase boundary.} Generation at which \(S_n\) first falls below half of \(3/(4\eta_s)\), from exact iteration of Eq.~\eqref{eq:survival22} along \(\eta_s=\eta_p\), plotted against the distance \(\epsilon\) from the boundary. The dashed line is the prediction of Eq.~\eqref{eq:ncross}, \(n_\times=(8\sqrt2\,\pi/9)\,\epsilon^{-1/2}\), with no fitted parameters. The points lie a constant two to three generations below it, so the relative deviation vanishes as \(\epsilon\to0\).}
    \label{fig:csd}
\end{figure}

\emph{Below the boundary} (\(\epsilon>0\)) the right-hand side is always negative, so \(S_n\to0\), but it is only of order \(\epsilon\) while \(|x_n|\lesssim\sqrt\epsilon\). \(S_n\) therefore lingers near \(3/(4\eta_s)\) before collapsing. The steps in this bottleneck are small, so the sum over generations can be replaced by an integral \cite{strogatz2015nonlinear}, giving the number of generations spent there,
\begin{equation}
    n_\times=\int_{-\infty}^{\infty}\frac{\mathrm{d}x}{\tfrac{81}{256}\epsilon+2x^2}=\frac{8\sqrt2\,\pi}{9}\,\epsilon^{-1/2}.
    \label{eq:ncross}
\end{equation}
The exponent \(1/2\) follows only from the quadratic shape of the map at the merging point, and is exact. The prefactor is exact as \(\epsilon\to0\); at finite \(\epsilon\), the time taken to reach and leave the bottleneck adds a correction of a few generations that does not grow with \(\epsilon^{-1/2}\). Figure~\ref{fig:csd} confirms this by exact iteration of Eq.~\eqref{eq:survival22}: the generation at which \(S_n\) first falls below half of \(3/(4\eta_s)\) lies two to three generations below \(n_\times\) for \(10^{-6}\le\epsilon\le10^{-1}\), and a fit for \(\epsilon\le3\times10^{-4}\) gives the exponent \(-0.501\).

\emph{Above the boundary} (\(\epsilon<0\)), Eq.~\eqref{eq:normal} has its stable fixed point at \(x=\tfrac{9}{16}\sqrt{|\epsilon|/2}\), where the slope of the map is \(1-\tfrac{9}{4\sqrt2}\sqrt{|\epsilon|}\). \(S_n\) therefore approaches \(S_\infty\) as \(e^{-n/\tau}\) with \(\tau\simeq(4\sqrt2/9)\,|\epsilon|^{-1/2}\), the same exponent. \emph{On the boundary} (\(\epsilon=0\)), \(x_{n+1}=x_n-2x_n^2\) gives the slower, algebraic approach \(S_n-S_\infty\simeq1/(2\eta_s n)\).

\paragraph{Finite networks.} Critical slowing down sets how large a finite network can be while retaining a substantial chance of survival below the boundary. The divergence of \(n_\times\) is mild in \(n\), but because the network size grows as \(4^n\), even a modest number of generations corresponds to a very large network. For example, at \(\eta_s=\eta_p=0.91\), less than \(1\%\) below the critical value \(54/59\) (\(\epsilon\approx0.013\), \(n_\times\approx35\)), \(S_n\) still exceeds \(3/4\) at \(n=23\), a network of \(4^{23}\approx7\times10^{13}\) links, and collapses only near \(n=35\). For perfect links, \(c_0=1\), \(S_n\) is also the probability of delivering a Bell pair. For imperfect links the survival window is useful only when \(c_0\) is close to one: with \(c_0=0.99\), a near-perfect connection (\(X_{20}\ge0.9\)) is still obtained with probability \(0.72\), whereas with \(c_0=0.85\) the mean concurrence has already fallen to \(\mathbb{E}[X_{10}]\approx0.01\) while \(S_{10}\) still exceeds \(3/4\): weaker links lose their strength long before they fail (main text, Fig.~5).

\section{Mean-field and higher-order closures for the surviving strength}
\label{sec:S3}

Write \(Y_n=\mathbb{E}[X_n\mid X_n>0]\) and \(p_n=\eta_sS_n^2\) as in Sec.~\ref{sec:S2}. A surviving branch has concurrence \(A=X_aX_b\), the product of two independent surviving subnetwork concurrences, so \(\mathbb{E}[A]=Y_n^2\) exactly. Counting the surviving outcomes of one motif as in Eq.~\eqref{eq:survival22} gives the exact relation
\begin{equation}
    S_{n+1}Y_{n+1}=2p_n(1-p_n)\,Y_n^2+\eta_p p_n^2\,\mathbb{E}\big[\hat P(A,B)\big],
    \label{eq:exactY}
\end{equation}
where \(A\) and \(B\) are the concurrences of the two surviving branches. The first term is exact. The second is not determined by \(Y_n\), because \(\hat P\) is nonlinear and \(\mathbb{E}[\hat P(A,B)]\neq\hat P(\mathbb{E}[A],\mathbb{E}[B])\).

\begin{figure}[!tb]
    \centering
    \includegraphics{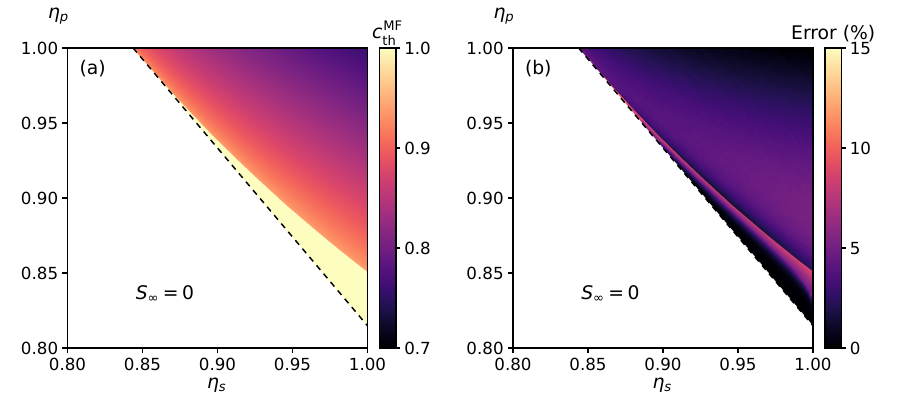}
    \caption{\textbf{Mean-field concurrence threshold.} (a)~Mean-field threshold \(c_{\mathrm{th}}^{\mathrm{MF}}\), the unstable fixed point of Eq.~\eqref{eq:MF} with \(S_n\) replaced by \(S_\infty\), on the same color scale as Fig.~4(b). The pale band beside the boundary, where \(c_{\mathrm{th}}^{\mathrm{MF}}=1\), is an artifact of the closure. (b)~Percentage deviation \(100\times|c_{\mathrm{th}}^{\mathrm{MF}}-c_{\mathrm{th}}|/c_{\mathrm{th}}\) from the threshold of Fig.~4(b). The error is a few percent over most of the surviving phase, vanishes toward \((\eta_s,\eta_p)=(1,1)\), and is largest, up to about \(13\%\), along a narrow ridge at the edge of the artifact band.}
    \label{fig:MF}
\end{figure}

\paragraph{Mean field.} Replacing each branch by its mean, \(\mathbb{E}[\hat P(A,B)]\to\hat P(Y_n^2,Y_n^2)\), closes the recursion:
\begin{equation}
    Y_{n+1}=\frac{2p_n(1-p_n)\,Y_n^2+\eta_p p_n^2\,\hat P(Y_n^2,Y_n^2)}{S_{n+1}},\qquad Y_0=c_0 .
    \label{eq:MF}
\end{equation}
For \(\eta_s=\eta_p=1\), \(S_n=p_n=1\) and Eq.~\eqref{eq:MF} reduces to the exact Eq.~(5). Replacing \(S_n\) by \(S_\infty\) gives a map with stable fixed points \(Y^*=0\) and \(Y^*=1\) and, for sufficiently reliable operations, an unstable fixed point that defines the mean-field threshold \(c_{\mathrm{th}}^{\mathrm{MF}}\) [Fig.~\ref{fig:MF}(a)]. Close to the boundary the unstable fixed point disappears and \(Y^*=1\) becomes unstable, so \(c_{\mathrm{th}}^{\mathrm{MF}}\) jumps to \(1\). This band is absent from the full numerics [Fig.~4(b)], and the mean-field error is largest along its edge [Fig.~\ref{fig:MF}(b)].

\paragraph{Second-order closure.} We also track \(Q_n=\mathbb{E}[X_n^2\mid X_n>0]\), so that a surviving branch has mean \(\mu_n=Y_n^2\) and variance \(\sigma_n^2=Q_n^2-Y_n^4\), both exactly. Expanding \(\hat P\) to second order about \(\mu_n\),
\begin{equation}
    \mathbb{E}\big[\hat P(A,B)\big]\simeq\hat P(\mu_n,\mu_n)+\tfrac12\sigma_n^2\big(\partial_1^2+\partial_2^2\big)\hat P(\mu_n,\mu_n),
\end{equation}
and likewise for \(\mathbb{E}[\hat P(A,B)^2]\). Inserting these into Eq.~\eqref{eq:exactY} and its analogue for the second moment,
\begin{equation}
    S_{n+1}Q_{n+1}=2p_n(1-p_n)\,Q_n^2+\eta_p p_n^2\,\mathbb{E}\big[\hat P(A,B)^2\big],
\end{equation}
gives a closed recursion for \((Y_n,Q_n)\), with \(Y_0=c_0\) and \(Q_0=c_0^2\).

\paragraph{Accuracy.} Table~\ref{tab:closures} compares both closures with the threshold obtained from the full distribution. To compare like with like, each closure is iterated from \(c_0\) for 50 generations, as in the numerics, rather than evaluated at its fixed point, and its threshold is located by bisection on \(c_0\). Far from the boundary (\(|\epsilon|\gtrsim0.08\)) both closures are then accurate to better than \(0.5\%\). Near the boundary both err by up to several percent, and retaining the variance does not remove the error. The distribution of \(X_n\) must therefore be propagated in full.

\begin{table}[!tb]
    \centering
    \caption{\textbf{Accuracy of the mean-field and second-order closures.} Concurrence threshold from the full distribution of \(X_n\) at \(n=50\), and the signed relative error \(100\times(c_{\mathrm{closure}}-c_{\mathrm{th}})/c_{\mathrm{th}}\) of each closure iterated to the same generation. The first two points lie far from the boundary; the last four lie close to it, \(|\epsilon|\le0.02\) (Sec.~\ref{sec:csd}).}
    \label{tab:closures}
    \begin{ruledtabular}
    \begin{tabular}{cccc}
        \((\eta_s,\eta_p)\) & \(c_{\mathrm{th}}\) & Mean field & Second order\\
        \hline
        \((0.99,\,0.99)\) & 0.771 & \(+0.12\%\) & \(-0.03\%\)\\
        \((0.95,\,0.95)\) & 0.821 & \(+0.20\%\) & \(-0.19\%\)\\
        \((0.90,\,0.9513)\) & 0.869 & \(-1.20\%\) & \(+3.86\%\)\\
        \((0.93,\,0.9164)\) & 0.893 & \(-3.18\%\) & \(+0.80\%\)\\
        \((0.90,\,0.9378)\) & 0.927 & \(-5.86\%\) & \(-2.11\%\)\\
        \((0.96,\,0.8814)\) & 0.935 & \(-6.37\%\) & \(-4.10\%\)\\
    \end{tabular}
    \end{ruledtabular}
\end{table}

\section{Ideal concurrence threshold}
\label{sec:S4}

For \(\eta_s=\eta_p=1\), Eq.~(5) reads \(c_{n+1}=\hat P(c_n^2,c_n^2)\). Below saturation, Eq.~\eqref{eq:P_rule} gives \(M=\lambda^2\) with \(\lambda=\lambda_1(c^2)\), so the nontrivial fixed point solves
\begin{equation}
    c=2\lambda\sqrt{1-\lambda^2},\qquad \lambda=\tfrac12\big(1+\sqrt{1-c^4}\big).
\end{equation}
Solving this single algebraic equation by bisection gives the threshold to any desired precision \cite{PhysRevLett.126.170501,doi:10.1126/sciadv.adt2404}, \(c_{\mathrm{th}}=0.759470997087017\ldots\). The slope of the map there is \(1.669>1\), so the fixed point is unstable and separates the basins of \(Y^*=0\) and \(Y^*=1\) [Fig.~4(a)].

\section{Numerical method}
\label{sec:S5}

At each generation we store the distribution of \(X_n\) as a probability mass function: the failure probability \(\mathbb{P}(X_n=0)=1-S_n\), and probabilities \(\pi_n(x_i)\) on the uniform grid \(x_i=i/(N_{\mathrm{bin}}-1)\). One renormalization step follows the network structure:
\begin{enumerate}
    \item \emph{Series.} For a branch of length \(U\), form the distribution of the product of \(U\) independent values drawn from \(\pi_n\) [Eq.~\eqref{eq:S_rule}] and multiply it by \(\eta_s^{U-1}\). The remaining mass is branch failure.
    \item \emph{Parallel.} Combine every pair of surviving values from the two branches through \(\hat P\) with weight \(\eta_p\). If only one branch survives, pass its value on unchanged. All other mass is added to \(\mathbb{P}(X_{n+1}=0)\).
    \item \emph{Quantization.} Round the new values onto the grid, discard entries with probability below \(\varepsilon_{\mathrm{prune}}\), and renormalize so that the total probability stays exactly one.
\end{enumerate}
Figure~\ref{fig:EXn} uses \(N_{\mathrm{bin}}=2001\). Figure~5 uses the same spacing, \(5\times10^{-4}\), refined geometrically (ratio \(1.005\)) below \(X=5\times10^{-3}\) and excluding \(X=0\), so that the weak surviving connections shown in its \(n=10\) insets are resolved and never rounded into the failure probability; the thresholds use \(N_{\mathrm{bin}}=1001\), \(\varepsilon_{\mathrm{prune}}=10^{-14}\), and \(n=50\) (\(L=2^{50}\)).

\paragraph{Threshold extraction.} At large \(n\), the probability of a near-perfect connection is close to a step function of \(c_0\). We take \(c_{\mathrm{th}}\) to be the smallest \(c_0\) for which the mass in the top four bins, \(M_{\mathrm{high}}(c_0)\), reaches half of its value for perfect links, \(M_{\mathrm{high}}(1)\), and locate it to within \(10^{-3}\) by bracketing and bisection on \(c_0\), propagating to \(n=50\) at each step. Points with \(S_\infty=0\), identified from the exact recursion, are excluded. Thresholds are computed on a \(101\times101\) grid in \((\eta_s,\eta_p)\); Fig.~4(b) shows a cubic interpolation of these values for display.

\section{Numerical accuracy and convergence}
\label{sec:S6}

The distribution is propagated deterministically, so there is no sampling noise. The remaining errors come from the grid, the pruning, and the finite generation.
\begin{itemize}
    \item The survival probability read from the numerics agrees with the exact Eq.~(2) to better than \(10^{-6}\) at over \(99\%\) of points in the surviving phase.
    \item The extracted \(c_{\mathrm{th}}\) changes by less than \(10^{-3}\) when \(N_{\mathrm{bin}}\) is varied over \(\{1001,2001,4001\}\) at representative points, and is stable for \(n>20\). We therefore quote thresholds to three significant figures.
    \item For \(\eta_s=\eta_p=1\) the distribution remains a single value at every generation, and the extracted threshold agrees with \(c_{\mathrm{th}}=0.75947\ldots\) (Sec.~\ref{sec:S4}) to within \(10^{-3}\).
    \item Accuracy is lower in a narrow strip just above the boundary, where \(S_n\) approaches \(S_\infty\) only slowly (Sec.~\ref{sec:csd}).
\end{itemize}

\begin{figure}[!tb]
    \centering
    \includegraphics{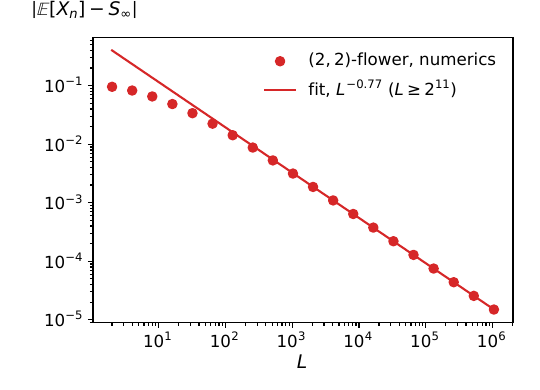}
    \caption{\textbf{Convergence of the mean end-to-end concurrence to the exact survival probability.} Deviation \(|\mathbb{E}[X_n]-S_\infty|\) of the numerically propagated mean from the exact fixed-point prediction, plotted against the shortest-path length \(L=2^n\) between A and B, for \(c_0=0.85\) and \(\eta_s=\eta_p=0.95\), above both thresholds (bottom row of Fig.~5). The straight line on double-logarithmic axes is a power-law fit for \(L\ge2^{11}\). As the distribution of \(X_n\) becomes binary, the surviving probability mass accumulates at \(X_n=1\), so \(\mathbb{E}[X_n]\to S_\infty=0.938218\), the value predicted by Eqs.~(2) and~(4).}
    \label{fig:EXn}
\end{figure}

Above both thresholds the distribution becomes binary, so \(Y_n\to1\) and \(\mathbb{E}[X_n]=S_nY_n\to S_\infty\) (main text, Fig.~5). Figure~\ref{fig:EXn} confirms this convergence up to \(L=2^{20}\approx10^6\): for \(L\ge2^{11}\) the deviation decays as a power law, \(|\mathbb{E}[X_n]-S_\infty|\propto L^{-0.77}\). Refining the grid from \(N_{\mathrm{bin}}=2001\) to \(8001\) changes these values by less than \(1\%\).

\end{document}